\documentclass[12pt]{article}

\begin{document}
\title{A Generalization of the Kodama State for Arbitrary Values of the Immirzi Parameter\footnote{An abridged version of
this paper was submitted to the Annual Essay Competition of the
Gravity Research Foundation, 2005}}
\author{Andrew Randono \\
               Center for Relativity, Department of Physics\\
        University of Texas\\
        Austin, TX 78712\\
        Email: arandono@physics.utexas.edu
}\date{} \maketitle
\begin{abstract}
The Kodama State for Lorentzian gravity presupposes a particular
value for the Immirzi-parameter, namely $\beta=-i$. However, the
derivation of black hole entropy in Loop Quantum Gravity suggests
that the Immirzi parameter is a fixed value whose magnitude is on
the order of unity but larger than one. Since the Kodama state has
de-Sitter spacetime as its classical limit, to get the proper
radiation temperature, the Kodama state should be extended to
incorporate a more physical value for $\beta$. Thus, we present an
extension of the Kodama state for arbitrary values of the Immirzi
parameter, $\beta$, that reduces to the ordinary Chern-Simons state
for the particular value $\beta=-i$. The state for real values of
$\beta$ is free of several of the outstanding problems that cast
doubts on the original Kodama state as a ground state for quantum
general relativity. We show that for real values of $\beta$, the
state is invariant under large gauge transformations, it is CPT
invariant (but not CP invariant), and it is expected to be
delta-function normalizable with respect to the kinematical inner
product. To aid in the construction, we first present a general
method for solving the Hamiltonian constraint for imaginary values
of $\beta$ that allows one to use the simpler self-dual and
anti-self-dual forms of the constraint as an intermediate step.
\end{abstract}
\section{Introduction}
The Kodama state\cite{Kodama:original, Kodama:original2} is the only
known exact solution to the constraints of quantum gravity which
also has a well-defined classical limit, namely de-Sitter
spacetime\cite{Smolin:kodamareview}. It is often overlooked that the
Kodama state presupposes a particular value for the Immirzi
parameter, namely $\beta=-i$. However, it has been shown that
requiring consistency with the entropy of isolated horizons from
Loop Quantum Gravity with the entropy of Hawking radiation fixes the
magnitude of the Immirzi parameter on the order of unity but larger
than one\cite{Ashtekar:entropy}. Furthermore, the Kodama state has
de-Sitter spacetime as its classical limit, and de-Sitter spacetime
is thermal due to the presence of an isolated cosmological horizon.
An analysis of the horizon entropy precisely parallels the
derivation of black hole entropy in Loop Quantum Gravity since in
both cases the entropy comes from boundary fields on an isolated
horizon. This again suggests that in order to arrive at the proper
horizon temperature the magnitude of the Immirzi parameter must be
larger than one. For these reasons alone it seems imperative to
extend the Kodama state to arbitrary values of the Immirzi
parameter.

In addition, the Kodama state is plagued with various difficulties
which have cast doubts on its viability as a ground state of quantum
general relativity: it is not invariant under large gauge
transformations, it is not normalizable under the physical inner
product, and it is not CPT invariant\cite{Witten:note}. All of these
difficulties can be traced down to the simple fact that there is no
$i$ in front of the Chern-Simons functional, which, in turn, can be
traced back to requirement that the Immirzi parameter is pure
imaginary. Thus, extending the state to real values of the parameter
has the potential of resolving some, if not all of these issues. We
show that for real values of $\beta$ when the familiar
pre-quantization condition for Chern-Simons theory is satisfied, the
extended state is invariant under large gauge transformations. In
addition, the state violates CP but is CPT invariant, and, drawing
from analogy with the Euclidean theory, it is expected to be
delta-function normalizable\cite{Smolin:linearkodama}.

To aid in the construction of the state we employ what seems to be a
general method for finding solutions to the Hamiltonian constraint
for arbitrary values of the Immirzi parameter which allows one to
use the simpler self-dual and anti-self dual forms as an
intermediate step. We present a general outline of the method below.

The equivalence of the equations of motion from the Einstein-Cartan
action and its (anti)self-dual counterpart is in part due to the
fact that the left and right handed components of the action are
non-interacting. Since they don't interact, often one can treat them
as independent variables despite one being the complex conjugate of
the other. In some ways this is analogous to complex Klein-Gordon
theory where $\phi$ and $\bar{\phi}$ are treated as functionally
independent variables. However, there are important differences. In
our case, the action is irreducibly complex and there is no analogue
of a $U(1)$ symmetry between the fields. The general idea is to
first treat the conjugate pair $\left(\omega_L, \Sigma_L \right)$
and $\left(\omega_R,\Sigma_R \right)$ as functionally independent
variables all the way up to the construction of the quantum
constraints and even solutions thereof, only then enforcing the
condition
\begin{equation}
{{\omega_R}^i}_j={{\overline{\omega_L}}^i}_j. \label{INDconstraint}
\end{equation}
As we will see, this allows one to work with the manageable
self-dual and anti-self-dual forms of the Hamiltonian constraint
while searching for solutions. Once a solution has been found,
enforcing an appropriate form the condition (\ref{INDconstraint}) is
simply a matter of making a change of variables in the solution and
the constraints. Written in terms of the new variables, barring
potential unforseen pitfalls, the new solution is automatically a
solution to the Hamiltonian constraint for the large class of
imaginary values of the Immirzi parameter subject to $|\beta|\geq
1$. Furthermore, for many solutions including the Kodama state, once
the extended solution has been found for these imaginary values of
$\beta$, nothing seems to prevent one from simply replacing $\beta$
with $i\beta$, thereby extending the solution to real values of the
parameter.

\section{Imaginary Immirzi Parameter}
Holst has shown\cite{Holst} that the constraint equations for
$(3+1)$ gravity in terms of the Barbero connection\cite{Barbero} for
an arbitrary value of the Immirzi parameter follow from a
modification of the Einstein-Cartan action of the form:
\begin{eqnarray}
S_H&=&S_{EC}+{S'}_H \nonumber\\
&=&\frac{1}{4k}\int_{M}\left(\epsilon_{IJKL}e^I\wedge e^J\wedge
\Omega^{KL} -\frac{2}{\beta}e^I\wedge e^J\wedge \Omega_{IJ}\right)
\label{HolstAction}
\end{eqnarray}
where $k=8\pi G$, $\beta$ is the Immirzi parameter, and
${\Omega^I}_J=d{\omega^I}_J+{\omega^I}_K\wedge{\omega^K}_J$ is the
curvature of a metric compatible (but not torsion free) spin
connection, ${\omega^I}_J$. Metric compatibility implies the gauge
group is either $SO(3,1)$ or, as we will assume in this section, its
covering group $SL(2,C)$.

For imaginary values of $\beta$, the Holst modification term has a
simple interpretation that will be essential for the rest of the
paper: it reflects a partial parity violation in the gravitational
interaction. To see this, we first write the Holst modified action
in a slightly more suggestive form. For convenience of notation, we
will work in the Dirac representation of the Lie algebra. Thus, take
the tetrad to be one-forms valued in the Dirac representation of the
Clifford algebra $e\equiv e^I\gamma_I$ and the connection
coefficients to be one-forms valued in the Lie algebra $sl(2,C)$,
$\omega\equiv \omega_{IJ}\frac{1}{4}\gamma^I\gamma^J$. Using the
identity
$Tr(\gamma^5\gamma^I\gamma^J\gamma^K\gamma^L)=4i\epsilon^{IJKL}$,
the action (\ref{HolstAction}) can then be written\footnote{Our sign
conventions are $\eta_{IJ}=diag(-1,1,1,1)$, $\epsilon^{0123}=-1$,
and $\gamma^5=i\gamma^0\gamma^1\gamma^2\gamma^3$.}
\begin{eqnarray}
S_H &=& \frac{1}{4k}\int_M Tr\left[-i\gamma^5e\wedge e\wedge \Omega
+\frac{1}{\beta}e\wedge
e\wedge\Omega\right] \nonumber\\
&=&\frac{-2i}{4k}\int_M
Tr\left[\left(\frac{1+(i/\beta)\gamma^5}{2}\right) \gamma^5 e\wedge
e \wedge\Omega \right].
\end{eqnarray}
The operator $\frac{1}{2}\left[1+(i/\beta)\gamma^5\right]$ is a
projection operator only for the particular values $\beta=-i$ and
$\beta=i$ where it projects the curvature into its left and right
handed components respectively, or, put another way, projects the
Lie algebra $sl(2,C)$ into $sl_\pm(2,C)$. We point out, however, for
imaginary values of $\beta$ such that $|\beta|>1$, the operator also
has a natural interpretation as the sum of two weighted chiral
projection operators. Motivated by partial parity violating
interactions we introduce coupling constants $\alpha_L=\cos^2\theta$
and $\alpha_R=\sin^2\theta$ where $0\leq\theta\leq\pi/2$. Then we
have,
\begin{equation}
\frac{\alpha_L}{2}\left(1-\gamma^5\right)+\frac{\alpha_R}{2}\left(1+\gamma^5\right)=\frac{1}{2}\left(1-\cos
2\theta\gamma^5\right).
\end{equation}
This simple observation has powerful consequences. It has long been
known that the Holst action reduces to the self-dual or
anti-self-dual action for the particular values $\beta=\mp i$. We
emphasize here, that this connection can be extended to a much
larger class of Immirzi parameter. With this goal in mind, using the
above construction, the natural generalization of the
Einstein-Cartan action to allow for partial-parity violation is
(inserting a factor of two for convenience):
\begin{eqnarray}
S &=& 2(\alpha_L S_{EC}[\omega_L]+\alpha_R S_{EC}[\omega_R])\nonumber\\
&=& \frac{-2i}{4k}\int_M Tr\left[ \alpha_L\ \gamma^5 e\wedge e\wedge
\Omega_{(L)} + \alpha_R\ \gamma^5 e\wedge e\wedge\Omega_{(R)} \right]\label{ppvaction}\\
&=&\frac{-2i}{4k}\int_M Tr\left[\left(\frac{1-\cos
2\theta\gamma^5}{2}\right)\gamma^5
e\wedge e \wedge\Omega \right]\nonumber\\
&=& \frac{1}{4k}\int_{M}\left[\epsilon_{IJKL}e^I\wedge e^J\wedge
\Omega^{KL} -2i\cos 2\theta \ e^I\wedge e^J\wedge
\Omega_{IJ}\right].\nonumber
\end{eqnarray}
Comparing this with the Holst-modified action we find that the
Immirzi parameter for this theory is given by
\begin{equation}
\beta=\frac{-i}{(\alpha_L-\alpha_R)}=\frac{-i}{\cos
2\theta},\label{ImaginaryImmirzi}
\end{equation}
and for these particular values, the physical meaning of the
parameter is clear: it is a numerical measure of the degree of
chiral asymmetry in the gravitational interaction\footnote{As a side
we note that chiral symmetry is violated in the standard model so
one might impose consistency relations which demand that the measure
of chiral asymmetry in the matter interactions of the standard model
is compatible with the measure of chiral violation in the
gravitational interaction (i.e. the Immirzi parameter). This could
yield an alternative fixing of the Immirzi parameter so long as one
accepts imaginary values for $\beta$. It is also worthy of
commentary that for these particular values, the Immirzi parameter
plays the dual role as the measure of chiral asymmetry and the
renormalization factor of Newton's constant, $G\rightarrow
G'=G/\beta$, as seen from the derivation of the entropy of thermal
radiation in the Kodama ground state\cite{Smolin:kodamareview}. This
interplay is intriguing, but it is not well understood.}. Thus, the
action (\ref{ppvaction}) is the bridge between the self-dual and
anti-self-dual actions, and the Holst action for the large class of
Immirzi parameter in the form of (\ref{ImaginaryImmirzi}). For this
reason, it is a natural starting point for the generalization of the
Kodama state to arbitrary values of the Immirzi parameter.

\section{Legendre Transforms}
Before attacking the action (\ref{ppvaction}) we take the time to
briefly review the Legendre Transform of the Holst Action. Our goal
here is to point out several subtleties that many sources ignore for
the sake of simplicity but play an important role in the
generalization of the Kodama state. Here it is easiest to work in a
vector representation. Since we eventually want to construct the
Kodama state, we also add in a cosmological constant term to the
action:
\begin{eqnarray}
S&=&S_{H}+S_{CC} \nonumber\\
&=&\frac{1}{4k}\int_{M}\left(\epsilon_{IJKL}e^I\wedge e^J\wedge
\left(\Omega^{KL}-\frac{\Lambda}{6}e^K\wedge e^L\right)
-\frac{2}{\beta}e^I\wedge e^J\wedge \Omega_{IJ}\right)
\label{HolstAction2}
\end{eqnarray}

Per usual, we decompose spacetime into spacelike foliations
$M=\Re\times\Sigma$ and introduce a timelike vector field
$\bar{t}=N\bar{n}+\bar{N}$ where $\bar{n}$ is the metric normal to
the foliation and $\bar{N}$ is metrically parallel to the foliation.
The canonical one form $dt$ associated with $\bar{t}$ defines the
time coordinate $t$. For the remainder of this section and the next
(with some obvious exceptions) all forms, duals, and lower case
Latin indices will be defined in the foliations $\Sigma$. We will
fix the gauge by assuming that the vector $\bar{n}$ is constant when
viewed as a vector in the vector bundle and has components
$n^I=e^I(\bar{n})=(1,0,0,0)$. Thus, the extrinsic curvature is
${K^i}=(q^\mu_aq^i_I\mathcal{D}_\mu n^I)dx^a={\omega^i}_0$, where
$q^i_I$ and $q^\mu_a$ project components in $TM$ to the foliations
$T\Sigma$. It is easiest to begin by decomposing the action into
components parallel and perpendicular to $\bar{n}$ and later making
the substitution $\bar{n}=\frac{\bar{t}-\bar{N}}{N}$. However you do
it the decomposition of (\ref{HolstAction}) can be written:
\begin{eqnarray}
\frac{-1}{k\beta}\int_{\Re\times\Sigma} dt &\wedge&[ E^i\wedge E^j
\wedge\pounds_{\bar{t}}A_{ij}-E^i\wedge E^j
\wedge \pounds_{\bar{N}}A_{ij} \nonumber\\
&-& A_{ij}(N\bar{n})G^{ij}-K_{ij}(N\bar{n})\tau^{ij}-N H ].
\end{eqnarray}
where $A_{ij}=\omega_{ij}-\beta K_{ij}$ is the Barbero connection
($K_{ij}\equiv \epsilon_{ijk}K^k$). We see in the first term the
familiar phase space consisting of the ``position" variable $A_{ij}$
and its conjugate momentum $P^{kl}=\frac{-1}{k\beta}E^k\wedge E^l$
whose Poisson brackets become operator commutators in the quantum
theory:
\begin{eqnarray}
\left[A_{ij}|_P,E^k\wedge
E^l|_Q\right]=-ik\beta\delta^k_{[i}\delta^l_{j]}\tilde{\delta}(P,Q)
\end{eqnarray}
where $\tilde{\delta}(P,Q)$ is the delta distribution valued 3-form
satisfying:
\begin{equation}
\int_{P\in\Sigma}f(P)\tilde{\delta}(P,Q)=f(Q).
\end{equation}

Here, the variables $N$, $\bar{N}$, $A_{ij}(N\bar{n})=N
A_{ij\mu}n^\mu$, and $K_{ij}(N\bar{n})=N\epsilon_{ijk}{K^k}_\mu
n^\mu$ are all Lagrange multipliers whose variation yields the
constraints below:

\textbf{Diffeomorphism Constraint:}
\begin{equation}
\pounds_{\bar{V}}A_{ij}\wedge E^i\wedge E^j\approx0 \label{diffeo1}
\end{equation}
for any smooth vector field $\bar{V}$ lying completely in $\Sigma$.
This constraint comes from varying the shift $\bar{N}$. The form of
the constraint differs trivially from the form usually presented in
the literature $E^i\wedge E^j\wedge F_{ij}(\bar{N})=E^i\wedge
E^j\wedge(\pounds_{\bar{N}}A_{ij}-D_A(A_{ij}(\bar{n})))$ because we
have removed the Gauss part of the constraint. This is accomplished
by taking $A_{ij}(N\bar{n})$ as a Lagrange multiplier as opposed to
$A_{ij}(\bar{t}).$

\textbf{Torsion Constraint}:
\begin{equation}
-2\beta {K^{[i}}_m \wedge E^m\wedge E^{j]}\approx 0.
\end{equation}
 This constraint comes from
varying $K^{ij}(N\bar{n})$. One can show that it is classically
equivalent to the vanishing of the normal component of the torsion
of the connection $\omega$ on the spacelike hypersurfaces: $n_I
D_\omega e^I=n_I T^I\approx 0$.

\textbf{Gauge Constraint}:
\begin{equation}
D_A(E^i\wedge E^j) +2\beta {K^{[i}}_m \wedge E^m\wedge
E^{j]}=D_{\omega}(E^i\wedge E^j)\approx 0.
\end{equation}
This constraint comes from varying the action with respect to
$A_{ij}(N\bar{n})$. It is the usual Gauss constraint with an extra
piece which is identical to the torsion constraint. It is
classically equivalent to the vanishing of the three dimensional
components of the torsion of the connection $\omega$:
$D_{\omega}E^i=T^i\approx 0$.

For imaginary values of $\beta$, the Gauge and the Torsion
constraints can be combined simply by adding them to yield the
ordinary form of the Gauss constraint:
\begin{equation}
D_A(E^i\wedge E^j)\approx 0.\label{Gauss1}
\end{equation}
This changes nothing since the real and imaginary parts must vanish
separately. For real values of $\beta$ one does not have this luxury
and one or the other constraint must be solved prior to
quantization. Typically one solves the gauge constraint by taking
$\omega=\omega[E]$ to be torsion free and then replacing the Torsion
and Gauge constraints by the single Gauss constraint (\ref{Gauss1}).

\textbf{Hamiltonian Constraint}:
\begin{equation}
\epsilon_{ijk}E^i\wedge\left(\Omega^{jk}_A+\frac{1}{\beta}(1+\beta^2)D_\omega
K^{jk}-(1+\beta^2){K^j}_m\wedge K^{mk}-\frac{\Lambda}{3}E^j\wedge
E^k\right)\approx 0. \label{Ham1}
\end{equation}
This comes from varying the lapse $N$. It has the usual form of the
Hamiltonian constraint with the addition of the term proportional to
$\epsilon_{ijk}E^i\wedge D_\omega K^{jk}$. This is usually ignored
because it vanishes identically when the both the Torsion and Gauge
constraints are applied. Keeping the extra term seems to complicate
the constraint unnecessarily, but as we will see it actually
simplifies our situation considerably because it facilitates the
connection between this Hamiltonian constraint and that of the
action (\ref{ppvaction}).

\section{Constraints from partial-parity violating action}
We begin by constructing the quantum constraints assuming $\omega_L$
and $\omega_R$ are functionally independent. Using a bit of
self-dualology, we can rewrite the action (\ref{ppvaction}) with a
cosmological constant:
\begin{eqnarray}
\frac{i\alpha_L}{k}\int_M\Sigma_{(L)IJ}\wedge\left(\Omega_{(L)}^{IJ}-\frac{\Lambda}{6}\Sigma_{(L)}^{IJ}
\right)-\frac{i\alpha_R}{k}\int_M
\Sigma_{(R)IJ}\wedge\left(\Omega_{(R)}^{IJ}-\frac{\Lambda}{6}\Sigma_{(R)}^{IJ}\right).
\end{eqnarray}
where
\begin{eqnarray*}
\Sigma_{(L/R)}^{IJ}=\frac{1}{2}\left(e^I\wedge e^J \mp
\frac{i}{2}{\epsilon^{IJ}}_{KL}e^K\wedge e^L\right).
\end{eqnarray*}
It is clear then that the action is essentially the difference of
two functionally independent actions, one self-dual and one
anti-self-dual, each of whose constraints are well known. As we will
see, this makes the constraints much easier to solve. We now
construct the Legendre transform of the above action. The technique
is the same as before. Although technically we do not have to fix
the gauge in the self-dual formalism, we do so as a matter of
convenience to compare the resulting constraints to those above. The
result is that the Legendre transformed action takes the form:
\begin{eqnarray}
S&=&\frac{-1}{k}\int_{\Re\times\Sigma}dt\wedge(i\alpha_L\Sigma_{(L)ij}\wedge\pounds_{\bar{t}}\omega_{(L)}^{ij}
-i\alpha_R \Sigma_{(R)ij}\wedge\pounds_{\bar{t}}\omega_{(R)}^{ij}\nonumber\\
&-&i\alpha_L\Sigma_{(L)ij}\wedge\pounds_{\bar{N}}\omega_{(L)}^{ij}
+i\alpha_R \Sigma_{(R)ij}\wedge\pounds_{\bar{N}}\omega_{(R)}^{ij}\nonumber\\
&-&i\alpha_L\omega_{(L)ij}(N\bar{n})D_{(L)}\Sigma_{(L)}^{ij}+i\alpha_R\omega_{(R)ij}(N\bar{n})D_{(R)}\Sigma_{(R)}^{ij}\nonumber\\
&-&NH).
\end{eqnarray}
The phase space now consists of the ``position" variables
$\omega_{(L)ij}$ and $\omega_{(R)ij}$ and their conjugate momenta
$P_{(L)}^{kl}=\frac{-i\alpha_L}{k}\Sigma_{(L)}^{kl}$,
$P_{(R)}^{kl}=\frac{i\alpha_R}{k}\Sigma_{(R)}^{kl}$ which yield the
quantum commutators\footnote{We apologize beforehand for being
rather loose about the quantization. The primary intent of this
paper is to construct the generalized Kodama state and outline a
method for solving the quantum constraints by means of the Kodama
state as an example. A more rigorous treatment of the quantization
is certainly in order.}:

\begin{eqnarray}
\left[\omega_{(L)ij}|_P,\Sigma_{(L)}^{kl}|_Q\right]&=&
\frac{-k}{\alpha_L}\delta^k_{[i}\delta^l_{j]}\tilde{\delta}(P,Q)\nonumber\\
\left[\omega_{(L)ij}|_P,\Sigma_{(R)}^{kl}|_Q\right]&=&0 \nonumber\\
\left[\omega_{(R)ij}|_P,\Sigma_{(R)}^{kl}|_Q\right]&=&
\frac{k}{\alpha_R}\delta^k_{[i}\delta^l_{j]}\tilde{\delta}(P,Q)\\
\left[\omega_{(R)ij}|_P,\Sigma_{(L)}^{kl}|_Q\right]&=&0
\label{Comm2}
\end{eqnarray}
These operators act on states in the expanded Hilbert space which we
take at the kinematical level to be $L^2[\Omega_{(L)}
\otimes\Omega_{(R)}]$ with respect to some presently undefined inner
product. Here $\Omega_{(L/R)}$ is the space of generalized
connections of $\omega_{(L/R)}$.

Now, $N$, $\bar{N}$, $\omega_{(L)ij}(N\bar{n})$, and
$\omega_{(R)ij}(N\bar{n})$ are to be treated as independent Lagrange
multipliers whose variation yields the constraints:
\begin{eqnarray}
D_{(L)}\Sigma_{(L)}^{ij} &\approx&  0 \nonumber\\
D_{(R)}\Sigma_{(R)}^{ij} &\approx&  0 \nonumber\\
\alpha_L\pounds_{\bar{V}}\omega_{(L)ij}\wedge\Sigma_{(L)}^{ij}
- \alpha_R\pounds_{\bar{V}}\omega_{(R)ij}\wedge\Sigma_{(R)}^{ij} &\approx& 0 \nonumber\\
\alpha_L\ast\Sigma_{(L)jk}\wedge(\Omega_{(L)}^{jk}-\frac{\Lambda}{3}\Sigma_{(L)}^{jk})
+\alpha_R\ast\Sigma_{(R)jk}\wedge(\Omega_{(R)}^{jk}-\frac{\Lambda}{3}\Sigma_{(R)}^{jk})
&\approx& 0. \label{Constraints2}
\end{eqnarray}
These constraints should be viewed as quantum operators (with the
given choice of operator ordering) acting on the Hilbert space.

In all of the above we have treated $\omega_{(L)}^{ij}$ and
$\omega_{(R)}^{ij}$ and their momenta as functionally independent
variables. We now need to impose the condition that
$\omega_{(R)}^{ij}=\overline{\omega}_{(L)}^{ij}$. Classically this
would simply entail a reduction of the phase space. However, we have
already quantized the theory, so the reduction is a reduction on the
Hilbert space itself. This is implemented by imposing operator
constraints on the configuration variables and momentum operators
and rewriting all the states in terms of the new variables. Since
the classical phase space for an arbitrary value of the Immirzi
parameter consists of the position-momentum pair $(A_{ij},-k\beta
E^m\wedge E^n)$ we expect that in the connection representation the
``position" operator will be the multiplicative operator $A^{ij}$
and the momentum operator will be $ik\beta\frac{\delta}{\delta
A_{ij}}$ for the particular value $\beta=-i/\cos 2\theta$. These
operators will act on states in the kinematical Hilbert space given
by $L^2[\mathcal{A}]$ where $\mathcal{A}$ is the space of
generalized connections of $A$. To this end, we \textit{define} the
``real" and ``imaginary" parts of $\omega_{(L)}$\footnote{The terms
real and imaginary are only heuristic labels here since these are
truly operators on the Hilbert space. Perhaps the terms Hermitian
and anti-Hermitian would be more appropriate, but we have not yet
defined the inner product or the Hermitian adjoint.}:
\begin{eqnarray}
\omega^{ij}\equiv Re(\omega_{(L)}^{ij})=\frac{1}{2}(\omega_{(L)}^{ij}+\omega_{(R)}^{ij})\nonumber\\
K^{ij}\equiv
Im(\omega_{(L)}^{ij})=\frac{1}{2i}(\omega_{(L)}^{ij}-\omega_{(R)}^{ij}),
\end{eqnarray}
and we define the new connection
\begin{equation}
A^{ij}\equiv
\omega^{ij}+\frac{i}{cos2\theta}K^{ij}=\frac{\alpha_L\omega_{(L)}^{ij}
-\alpha_R\omega_{(R)}^{ij}}{\alpha_L-\alpha_R}.
\end{equation}
In addition, we define a surface form, which as we will see is
proportional to the conjugate momentum of $A$:\footnote{The last
equality only holds because of our choice of gauge fixing. Without
this gauge fixing there would be extra terms.}
\begin{equation}
\Sigma^{ij}\equiv
\alpha_L\Sigma_{(L)}^{ij}+\alpha_R\Sigma_{(R)}^{ij}=E^i\wedge E^j.
\end{equation}
With these definitions, from the commutation relations (\ref{Comm2})
one deduces commutation relations for the new variables:
\begin{eqnarray}
\left[\omega_{ij}|_P,\Sigma^{kl}|_Q\right]&=&0 \nonumber\\
\left[K_{ij}|_P,\Sigma^{kl}|_Q\right]&=&
ik\delta^k_{[i}\delta^l_{j]}\tilde{\delta}(P,Q)\nonumber\\
\left[A_{ij}|_P,\Sigma^{kl}|_Q\right]&=& \frac{-k}{\cos
2\theta}\delta^k_{[i}\delta^l_{j]}\tilde{\delta}(P,Q)\label{Comm3}
\end{eqnarray}
These operators now act on the reduced Hilbert space. We note the
that vanishing of the commutator $\left[\omega,\Sigma\right]$
follows from the commutation relations (\ref{Comm2}) and does not
require that $\omega$ is torsion free, which is effectively imposed
by the Gauss constraint in the complex theory.

If we now make these substitutions in the constraints
(\ref{Constraints2}), after a bit of manipulation, the full set of
constraints reduces to the set (\ref{diffeo1})-(\ref{Ham1}) for the
class of Immirzi parameters given by $\beta=-i/\cos 2\theta$. This,
incidentally, is the reason why we took the time to work through the
constraints explicitly: the extra term in the Hamiltonian makes its
appearance in this redefinition of variables. Thus, keeping this
term as opposed to eliminating it by appeals to the Gauss constraint
makes more transparent the connection between the simple Hamiltonian
constraint of (\ref{ppvaction}) and that of the Holst action. This
suggests the following procedure for solving the constraint
equations: first work with the simpler form of the constraints
(\ref{Constraints2}) in order to find solutions. Then, once a
solution has been found, impose the condition
$\omega_{(R)}^{ij}=\overline{\omega}_{(L)}^{ij}$ by rewriting the
solution and the constraints in terms of the new variable $A$, whose
real and imaginary parts, $\omega$ and $K$, might separate in the
final form of the solution. The new solution should be a solution to
the full set of constraints for $\beta=-i/\cos 2\theta$. In the next
section, we demonstrate this explicitly for the generalization of
the Kodama state.

\section{The Generalized Kodama State}
The quantum constraints (\ref{Constraints2}) immediately admit the
Kodama-like solution
\begin{equation}
\Psi[\omega_{(L)},\omega_{(R)},\alpha_L,\alpha_R]=\exp\left[\frac{3}{2k\Lambda}\left(\alpha_L\int_\Sigma
Y_{CS}[\omega_{(L)}]-\alpha_R\int_\Sigma
Y_{CS}[\omega_{(R)}]\right)\right]
\end{equation}
where
\begin{eqnarray*}
\int Y_{CS}[\omega]&=&\int Tr\left(\omega\wedge
d\omega+\frac{2}{3}\omega\wedge\omega\wedge\omega\right)\\
&=&\int\left({\omega^i}_k\wedge
d{\omega^k}_i+\frac{2}{3}{\omega^i}_m\wedge {\omega^m}_n\wedge
{\omega^n}_i\right).
\end{eqnarray*}
We note that for the particular value $\alpha_L=1$ this reduces to
the original Kodama state. We also note that a strikingly similar
state with $\alpha_{L}=\alpha_{R}$ and an overall factor of $i$ was
found in the context of quantum supergravity in
\cite{Ying:supergravity}.

Our task now is to rewrite
$\Psi[\omega_{(L)},\omega_{(R)},\alpha_L,\alpha_R]$ as an explicit
function of $A$, $K$, and $\beta$ and check that it is in fact a
solution to the constraints (\ref{diffeo1})-(\ref{Ham1}) for
$\beta=-i/(\alpha_L-\alpha_R)$. Using the shift identity,
\begin{equation}
\int Y_{CS}[\omega+\kappa]=\int \left(Y_{CS}[\omega]+Tr(\kappa\wedge
D_\omega \kappa+\frac{2}{3}\kappa\wedge \kappa\wedge \kappa)\right)
\nonumber
\end{equation}
one can show after a bit of algebra that the new state is,
\begin{eqnarray}
& & \Psi[\omega_{(L)},\omega_{(R)},\alpha_L,\alpha_R]\Rightarrow
\Psi[A,K,\beta]\\
&=&\exp\left[\frac{-3i}{2k\Lambda\beta}\int_{\Sigma}\left(Y_{CS}[A]-(1+\beta^2)Tr\left(K\wedge
D_\omega K -\frac{2}{3}\beta K\wedge K\wedge
K\right)\right)\right].\nonumber\label{Kodama}
\end{eqnarray}
To check that this is a solution to the Hamiltonian constraint
(\ref{Ham1}) for $\beta=-i/\cos2\theta$, we write the surface
operator in the connection representation\footnote{Here we are using
MTW's notation where $[abc]$ is simply the pure alternating symbol
to distinguish from the densitized alternating symbol
$\epsilon_{abc}\equiv \sqrt{|g|}[abc]$. In this expression, the
density is implicitly contained in the operator $\delta/\delta
{A^{ij}}_c$, which is proportional to the densitized triad
operator.}
\begin{equation}
E_i\wedge E_j=ik\beta\frac{\delta}{\delta A^{ij}}\equiv
ik\beta\left[abc\right] dx^a\wedge dx^b \frac{\delta}{\delta
{A^{ij}}_c}
\end{equation}
and the extrinsic curvature
\begin{equation}
K^{ij}=-\frac{A^{ij}-\omega^{ij}}{\beta}.
\end{equation}
Some simple arithmetic shows that $\psi[A,K,\beta]$ is in fact in
the kernel of the operator
\begin{equation}
\Omega^{jk}_A+\frac{1}{\beta}(1+\beta^2)D_\omega
K^{jk}-(1+\beta^2){K^j}_m\wedge K^{mk}-\frac{\Lambda}{3}E^j\wedge
E^k
\end{equation}
and therefore satisfies the Hamiltonian constraint for an
appropriate choice of operator ordering. It should be clear that the
state is invariant under (small) gauge transformations and
infinitesimal diffeomorphisms so we won't check that it satisfies
the diffeomorphism and and Gauss constraint explicitly.

\section{Extension to real values of the Immirzi Parameter}
Although our derivation of the generalized Kodama state technically
only holds for $\beta=-i/\cos 2\theta$, the reader may have already
noticed that the state is in fact in the kernel of the Hamiltonian
for any finite value of $\beta$, real or complex\footnote{The only
subtlety is that for real values of $\beta$ the Gauge constraint
must be solved so $\omega$ is torsion free and is explicitly a
function of derivatives of the triad.}. However, the state has very
different functional properties when the Immirzi parameter is real.
These new properties may resolve some of the outstanding issues
associated with the Kodama state as a valid ground state for general
relativity. However, we begin by briefly discussing in general how
to extend the procedure used in this paper to real values of the
Immirzi parameter.

Once one has obtained a solution to the simpler self-dual form of
the Hamiltonian constraint of (\ref{Constraints2}) and rewritten the
solution in terms of $A$, $K$, and $\beta=-i/(\alpha_L-\alpha_R)$,
one can then check whether the solution is still valid for arbitrary
values of $\beta$. Since the Hamiltonian constraint takes the same
form for any value of $\beta$, the solution $\psi[A,K,\beta]$ will
still be a solution to the Hamiltonian constraint \textit{so long as
$\psi[A,K,\beta]$ is functionally well-behaved after the
substitution} $\beta\rightarrow -i\beta$. Thus, we have reduced the
problem to a problem analogous to the following: given a complex
valued function $f(x,y)$ of two \textit{real} variables $x$ and $y$,
is the function still well-behaved when its domain is extended to
the complex plane of $y$? We are not guaranteed that our solution
will still be a solution when we make the substitution
$\beta\rightarrow i\beta$, but so long as the function is not
pathological upon a complex extension of $\beta$, the solution will
still hold. However, there are other requirements for a physical
solution which are dramatically affected by such an extension, for
example, normalizability and CPT invariance. As a simple, but
relevant example, the two functions $f(k,x)=e^{k x}$ and
$f(K=ik,x)=e^{K x}$ satisfy equations of the same form (think of
this as the Hamiltonian constraint): $\partial_x f(k,x)-kf(k,x)=0$
and $\partial_x f(K,x)-K f(K,x)=0$. However, one is delta-function
normalizable and CPT invariant, while the other is not. Thus,
substituting $\beta \rightarrow i\beta$ could make a physical state
unphysical or vice-versa. Indeed, this is true for the
generalization of the Kodama state to real values of $\beta$.

We note that, for real values, because of the $i$ in front of the
Chern-Simons functional, the state is invariant under large gauge
transformations so long as
$\kappa=\frac{3}{4G\Lambda\beta}=\frac{a_H}{4\beta a_{0}}$ is an
integer where $a_H=12\pi/\Lambda$ is the area of the cosmological
horizon of de-Sitter spacetime and $a_0$ is the area of a sphere
with a radius of the Planck length. This is the familiar
pre-quantization condition of Chern-Simons theory. This is related
to the analogous pre-quantization condition encountered in the
analysis of the surface degrees of freedom on an isolated horizon,
which also carries a Chern-Simons field. In the context of isolated
horizons, the condition is related to the integer number of
punctures on the horizon and the quantized deficit angles of the
connection around these punctures. We expect the same interpretation
in our context if the state is to reproduce some variant of
de-Sitter spacetime with a cosmological horizon.

In addition, the state is CPT invariant for real values of $\beta$.
Here we follow Soo's definitions of the action of discrete
symmetries\cite{Soo:CPT}. Since a graviton is its own anti-particle,
$C$ acts trivially on the state. Under parity $K^i=(q^\mu_a
q^i_I\mathcal{D}_\mu n^I)dx^a\rightarrow -K^i$. Since
${\epsilon^i}_{jk}\rightarrow {\epsilon^i}_{jk}$ (this is consistent
with $P$ \textit{actively} inverting the volume form) we have
${K^i}_j\rightarrow -{K^i}_j$.\footnote{This is also seen more
readily by noting that for $\beta=-i$,
$\omega^{ij}_{(L)}=\omega^{ij}+iK^{ij}$ is the pullback of the
left-handed four dimensional spin connection, $\omega^{IJ}_{(L)}$.
Since under parity left-handed spinors go to right-handed spinors,
the left-handed spin connection \textit{and} its pullback must also
go their right-handed counterparts. Thus, we deduce under parity
$K^{ij}\rightarrow -K^{ij}$.} But under time reversal we also have
${K^i}_j \rightarrow -{K^i}_j$ since, in our gauge, $n^I=t^I/N
\rightarrow -n^I$. Thus under $PT$, ${K^i}_j\rightarrow {K^i}_j$.
The connection coefficient $\omega$ is unaffected by $P$ and $T$.
Now, $T$ is anti-unitary so it also acts on the state by complex
conjugation introducing an overall minus sign to the phase. However,
the whole the Chern Simons functional is parity odd, and the terms
involving the extrinsic curvature have the same \textit{overall}
behavior under parity (ignoring the effect of $P$ on ${K^i}_j$ which
is canceled by the action of $T$). Thus, in total, the state
violates $CP$, but it is $CPT$ invariant.

Finally, we notice that for real values of $\beta$ the state is pure
phase. Thus, drawing from analogy with Kodama state in the Euclidean
theory, we expect that the state is delta-function normalizable with
respect to the kinematical inner-product\cite{Smolin:linearkodama}.

\section{Concluding Remarks}
We have shown that the Kodama state can be naturally extended to
arbitrary values of the Immirzi parameter. The result reduces to the
original Kodama state when the Immirzi parameter is $\beta=-i$.
Along the way we have employed what appears to be a general method
of solving the Hamiltonian constraint for arbitrary values of the
Immirzi parameter which allows one to use the simpler self-dual and
anti-self dual forms of the constraint as an intermediate step. This
method capitalizes on the interpretation of a large class of the
Immirzi parameter as the measure of chiral asymmetry in the
gravitational interaction. Much work is left to be done. First, it
would be valuable to know exactly what conditions must be satisfied
for our method of solution to work. Furthermore, a rigorous
extension of the method to spin network states is in order. Second,
an exploration of the consequences of $\beta=-i/\cos 2\theta$ as the
measure of partial parity violation in gravity may yield interesting
results. As alluded to in a footnote, this could lead to an
alternative fixing of the Immirzi parameter. Finally, much work is
left to be done on the generalized Kodama state itself. One needs to
show that it has a well-defined classical limit which reproduces
something like de-Sitter spacetime. The question of normalizability
needs to be dealt with rigorously. One may also question what effect
the generalization has on the loop transform of the Kodama state
which we know has a simple and elegant connection with knot
invariants\cite{Witten:knots} This is currently under investigation.

\section*{Acknowledgments}
This paper is dedicated to my former undergraduate advisor,
Allen Everett, now Professor Emeritus at Tufts University. I would
also like to thank Lee Smolin and Richard Matzner for their help in
finalizing this draft.

\bibliographystyle{utphys}
\bibliography{GeneralizedKodama}

\providecommand{\href}[2]{#2}\begingroup\raggedright\begin{thebibliography}{10}

\bibitem{Kodama:original}
H.~Kodama {\em Prog. Theor. Phys.} {\bf 80} (1988) 1024.

\bibitem{Kodama:original2}
H.~Kodama {\em Phys. Rev. D} {\bf 42} (1990) 2548.

\bibitem{Smolin:kodamareview}
L.~Smolin, ``Quantum gravity with a positive cosmological constant,''
  \href{http://www.arXiv.org/abs/arXiv:hep-th/0209079}{{\tt
  arXiv:hep-th/0209079}}.

\bibitem{Ashtekar:entropy}
A.~Ashtekar, J.~C. Baez, and K.~Krasnov, ``Quantum geometry of isolated
  horizons and black hole entropy,'' {\em Adv. Theor. Math. Phys.} {\bf 4}
  (2000) 1--94, \href{http://www.arXiv.org/abs/arXiv:gr-qc/0005126}{{\tt
  arXiv:gr-qc/0005126}}.

\bibitem{Witten:note}
E.~Witten, ``A note on the chern-simons and kodama wavefunctions,''
  \href{http://www.arXiv.org/abs/arXiv:gr-qc/0306083}{{\tt
  arXiv:gr-qc/0306083}}.

\bibitem{Smolin:linearkodama}
L.~Friedel and L.~Smolin {\em Class. Quant. Grav.} {\bf 21} (2004) 3831--3844,
  \href{http://www.arXiv.org/abs/arXiv:hep-th/0310224}{{\tt
  arXiv:hep-th/0310224}}.

\bibitem{Holst}
S.~Holst, ``Babero's hamiltonian derived from a generalized hilbert-palatini
  action,'' {\em Phys.Rev. D} {\bf 53} (1996) 5966--5969,
  \href{http://www.arXiv.org/abs/arXiv:gr-qc/9511026}{{\tt
  arXiv:gr-qc/9511026}}.

\bibitem{Barbero}
J.~F. Barbero, ``Real ashtekar variables for lorentzian signature
  space-times,'' {\em Phys. Rev. D} {\bf 51} no.~10, 5507--5510,
  \href{http://www.arXiv.org/abs/arXiv:gr-qc/9410014}{{\tt
  arXiv:gr-qc/9410014}}.

\bibitem{Ying:supergravity}
Y.~Ling and L.~Smolin, ``A holographic formulation of quantum supergravity,''
  {\em Phys.Rev. D} {\bf 63} (2001)
  \href{http://www.arXiv.org/abs/arXiv:hep-th/0009018}{{\tt
  arXiv:hep-th/0009018}}.

\bibitem{Soo:CPT}
C.~Soo, ``Self-dual variables, positive semi-definite action, and discrete
  transformations in four-dimensional quantum gravity,'' {\em Phys. Rev. D}
  {\bf 52} (1995) 3484--3493,
  \href{http://www.arXiv.org/abs/arXiv:gr-qc/9504042}{{\tt
  arXiv:gr-qc/9504042}}.

\bibitem{Witten:knots}
E.~Witten, ``Quantum field theory and the jones polynomial,'' {\em Commun.
  Math. Phys.} {\bf 121} (1989) 351--399.

\end{thebibliography}\endgroup

\end{document}